\documentclass[runningheads]{llncs}
\usepackage{lineno}
\usepackage[T1]{fontenc}
\usepackage{graphicx}
\usepackage{microtype}
\usepackage{blindtext}
\usepackage{hyperref}
\usepackage{xurl}

\begin{document}
\title{More Expert-like Eye Gaze Movement Patterns are Related to Better X-ray Reading}
\titlerunning{Eye Gaze and X-ray Reading Performance}
%
\authorrunning{P. Yang et al.}

\author{}
\institute{}
\author{Pingjing Yang\inst{1}\orcidID{0009-0000-3297-6030} \and
Jennifer Cromley\inst{1}\orcidID{0000-0002-6479-9080} \and
Jana Diesner\inst{1,2}\orcidID{0000-0001-8183-7109}}
\institute{University of Illinois Urbana-Champaign, Champaign IL 61820, USA \\
\email{\{py2,jcromley\}@illinois.edu.com}\\ \and
Technical University of Munich, Munich, Germany\\
\email{jana.diesner@tum.de}}

\maketitle         
\vspace*{-8mm}

\begin{abstract}
Understanding how novices acquire and hone visual search skills is crucial for developing and optimizing training methods across domains. Network analysis methods can be used to analyze graph representations of visual expertise. This study investigates the relationship between eye-gaze movements and learning outcomes among undergraduate dentistry students who were diagnosing dental radiographs over multiple semesters. We use network analysis techniques to model eye-gaze scanpaths as directed graphs and examine changes in network metrics over time. Using time series clustering on each metric, we identify distinct patterns of visual search strategies and explore their association with students’ diagnostic performance. Our findings suggest that the network metric of transition entropy is negatively correlated with performance scores, while the number of nodes and edges as well as average PageRank are positively correlated with performance scores. Changes in network metrics for individual students over time suggest a developmental shift from intermediate to expert-level processing. These insights contribute to understanding expertise acquisition in visual tasks and can inform the design of AI-assisted learning interventions.

\keywords{Eye-gaze Movement  \and Time Series Clustering \and Network Analysis.}
\end{abstract}

\section{Introduction}

Eye tracking is an established method for studying the allocation of experts' and novices' visual attention in a wide range of domains, e.g., medicine~\cite{ashraf2018eye}, sports~\cite{silva2022differences}, transportation~\cite{ziv2016gaze}, and education~\cite{grub2022professional}. At the same time, datasets capturing longitudinal eye tracking data that allow for studying how expertise develops over time are rare, which limits our knowledge about changes in eye-gaze patterns and related learning outcomes. Eye tracking studies often rely on aggregated metrics such as time spent in each Area of Interest (AOI) ~\cite{mason2013fourth} or the number of fixations per AOI~\cite{leng2024train}. Some studies have also applied network analysis methods to study characteristics of the entire scan path and their relationship with various outcomes, such as performance in flight training simulations~\cite{hua2022effect}.

The network metric of the number of nodes, also known as network size, quantifies the number of Areas of Interest (AOIs) fixated on. When applied to eye tracking scanpaths, network size is an indicator of the cumulative number of AOIs that a student has visited. For example, network size is higher for dermatologists trained to do more thorough examinations for skin cancer~\cite{dhengre2024investigating}.

The number of edges quantifies the number of saccades (moves) from any AOI to another AOI. When applied to eye tracking, the number of edges indicates the number of moves/transitions/switches that a student has made among AOIs. More transitions between mathematics questions and the representations needed to answer them are associated with higher test scores~\cite{zhu2015exploratory}.

The network diameter quantifies the longest yet most graph-efficient distance jumped across AOIs in the scanpath. When applied to eye tracking, the diameter represents the total distance in AOI ‘jumps’ across the entire stimulus set. The network diameter is significantly positively related to growth in test scores on written answers over time~\cite{ifenthaler2011mystery}.

Degree centrality quantifies the number of edges (saccade) per node (AOI) across all AOIs. When applied to eye tracking, degree centrality indicates how many pieces of visual information (represented as nodes), e.g.,  AOIs, a node is directly linked to. Higher degree centrality in eye tracking has been associated with better student performance~\cite{davalos2023identifying}.

Eigenvector centrality is a recursive function of degree centrality, thereby indicating the influence of an AOI by considering the degree centrality of a node's neighbors (in our case, other AOIs, in network terms also called alters, that are directly connected to an ego AI). Higher eigencentrality in eye tracking has been associated with better task performance on a web search task~\cite{loyola2014characterizing}.

Stationary entropy in networks measures the amount of uncertainty in the spatial distribution of a sequence of fixations. This metric is correlated with expertise in educational development: ~\cite{chanijani2016entropy} discovered that stationary gaze entropy (SGE) decreases as students’ expertise increases from novice to intermediate and from intermediate to expert in problem solving in physics. Network transition entropy measures the unpredictability of visual scanning patterns. Higher transition entropy suggests a more random pattern of scanning behaviors: ~\cite{schieber2008visual} found that gaze transition entropy (GTE) is lower for older drivers when completing a subsidiary loading task while driving on a two-lane rural highway. The mentioned studies suggest that entropy-based measures may be relevant for assessing expertise levels in visually demanding tasks.

In this paper, we apply network analysis metrics to an eye tracking data set collected from undergraduate dental students who visually inspected dental X-rays over a semester (with some participating for multiple semesters) and investigate the relationship between characteristics of the scanpath and students' performance on a dental anomaly detection task. 


\vspace{-4mm}

\section{Background}

\subsection{Analyzing Developmental Data}

Studying the development of learning—including its component skills, and other learning-related variables such as different aspects of motivation—is usually done by using growth curve modeling, which considers the shape of change in learning-related variables (increasing, decreasing, linear, curvilinear, discontinuous) and predictors of students’ initial scores and rates of change~\cite{singer2003applied}. The scores whose change is modeled over time might represent reading comprehension~\cite{pianta2008classroom}, answers to interest questionnaires~\cite{rotgans2017interest}, or other typical summed scales. Researchers have used this approach to analyze changes in physical (e.g., a child’s height; ~\cite{black1999predicting}) or physiological (e.g., heart rate; ~\cite{hostinar2014cortisol}) variables. In the present work, we model the change in selected network analytic metrics that capture aspects of a learner’s eye tracking scanpaths over time. 


We also hypothesize subsets of students with different patterns of change~\cite{pianta2008classroom} via grouping or clustering methods. These learning patterns are often modeled using a specialized type of growth curve modeling called growth mixture modeling~\cite{wickrama2021higher}. For example, one subset of students might show positive linear growth on a variable, and a different subset of students might show positive quadratic growth on the same variable, even though these two groups at the final testing point might have (statistically) similar or different scores. Different patterns of change can suggest educational implications, such as fostering a more adaptive trajectory of change or helping instructors understand that different trajectories can result in the same adaptive outcomes, depending on the findings. In other studies, differing growth curves might relate to different student clusters, which indicates that eye-gaze patterns correlate with differences in learning outcomes. In the present research, we test for different trajectories of change in network analysis metrics, which capture aspects of a learner’s eye tracking scanpaths.
\vspace{-3mm}

\subsection{Applying Network Analysis to Developmental Data}

Despite its limited application in educational research, network analysis has been used to quantify behavioral patterns. ~\cite{zhu2015exploratory} found no significant differences in network density or reciprocity between high- and low-performing math-solving groups but identified three triadic structures (003, 021D, and 111U) that significantly differed between the groups. More recently, ~\cite{ma2023eye} demonstrated that network metrics—including density, centrality, small-worldness, transitivity, and global efficiency—differentiated scanpath networks of low- and high-ability readers. On the other hand, current studies primarily employ network analysis for between-subject comparisons, using statistical methods such as t-tests ~\cite{zhu2015exploratory,ma2023eye} to establish correlations between network metrics and learning performance. In the present study, we synthesize existing research to evaluate the predictive power of network metrics in X-ray reading performance. This work advances education performance analysis by incorporating time-series modeling to capture dynamic learning processes.

\vspace{-4mm}

\section{Research Questions}

Based on our review of network metrics applied to visual processing, with a focus on expertise development and considering the context of undergraduate dentistry students looking for anomalies in dental X-rays over multiple semesters, we address the following questions: 
\vspace{-2mm}
\begin{enumerate}
 \item What can we learn about expertise development from the shape of change in various network analysis metrics applied to the students’ scanpaths data over time?

 \item Are there subsets or clusters of students who show different patterns of change (i.e., different developmental routes)? If so, how do these clusters relate to student performance?

 \item What do the patterns of change in network analysis metrics suggest about the novice-intermediate-expert developmental continuum? 
\end{enumerate}
\vspace{-2mm}
This research advances the generally available and validated set of methods and metrics for studying the impact of visual attention allocation on learning outcomes. It also improves our understanding of the potential capabilities of AI technologies for analyzing X-rays: Humans reading X-rays is an instance of a qualitative analysis method applied to qualitative data; a task that requires substantial and domain-specific training to lead to reliable results \cite{bernard2016analyzing}. How good can AIs be at this task? If human learners advance in their ability to interpret X-rays with continued training and achieve more correct results over time, then using such data to train AI models for X-ray assessment has a chance to result in models that lead to potentially reliable results. If, however, training humans does not lead to sustained improvements in their ability to read X-rays correctly, then we should not assume that AIs such as large vision models have or develop this ability after being trained on any scale of prior data. Our paper also sheds light on this question.   

\vspace{-4mm}

\section{Methods}

\subsection{Data}

The dataset we use was collected and made publicly available by a research team headed by Fabian Huettig and Constanze Keutel~\cite{borchers2023time,castner2022gaze} with funding from the Leibniz-WissenschaftsCampus program Cognitive Interfaces. We downloaded the data from PsychArchives from \url{http://dx.doi.org/10.23668/psycharchives.5681} and additional code from \url{https://github.com/conradborchers/visualsearchopt}. Additional clarification was provided by Conrad Borchers (personal communication on June 12 and August 26, 2024).
\vspace{-3mm}

\subsection{Participants and Educational Context}

Participants were 107 undergraduate dentistry students from the University of Tuebingen in Germany in 2017-2018. They took part in 90-minute-long session(s) during one or more semesters as early as their 6th semester (middle of 3rd year in the degree program) to as late as their 10th (graduating) semester in the program. At their first time of participation, their average age was 25.25 and they were 64\% female. They were paid 15 Euros for each eye tracking session that they participated in~\cite{borchers2023time}.

As part of their regular sequence of courses, all students in this program took a 6th-semester course in radiology that involved learning how to read dental X-rays and also included practice reading 100 dental X-rays that did not include the stimuli used in the dataset we analyzed. The general approach to teaching reading of X-rays is top-left to top right (on the X-ray), then bottom right to bottom left~\cite{cosson2020interpreting}.  Students might be considered novices at the beginning of the research study~\cite{eder2021support}. They might be considered intermediates when they had at least five previous semesters of learning about dental anatomy and dental health, including photographic images of healthy gums and teeth, as well as images of gum disease or dental caries, and were taking- or had taken- a radiology course.
\vspace{-3mm}

\subsection{Stimuli Presented to Participants and Performance Task}

In each session, students saw a series of ten dental X-rays, were asked to look for dental anomalies in each X-ray, such as evidence of gum disease or dental caries (e.g., cavities), and were eye tracked while examining the X-rays. Students were later asked to use their computer mouse to circle each anomaly in turn. In this paper, we analyze the anomaly detection scores provided in the dataset, and do not use the eye tracking from the marking session. Each anomaly was pre-defined as an AOI in the dataset. Since each X-ray might feature a different number of AOIs, we normalized the performance scores (e.g., finding 50\% of the possible anomalies in a particular X-ray) for each X-ray.
\vspace{-3mm}

\subsection{Eye Tracking Methods, Equipment, and Data Collection}

The SensoMotoric Instruments RED 250 eye tracker and SMR BeGaze software were are described in detail in the team’s publications~\cite{borchers2023time,castner2022gaze,richter2020massed,eder2021comparing,eder2022see,eder2021support}. After calibration on the mobile eye tracking equipment, each participant’s eye gazes were recorded for the duration of the X-ray task session. Data were excluded by the original research team if the tracking rate was below 80\%. 
\vspace{-3mm}

\subsection{Procedures}

After participants provided informed consent, they were seated in front of a computer, put on a pair of eye tracking glasses, and completed the manufacturer’s calibration check. They received instructions on how to mark any anomalies they detected. They were then shown the X-rays one at a time, with a 90-second free-exploration phase and a subsequent untimed marking anomalies phase. Subsequently, participants completed a dental conceptual knowledge test and a demographics form, which are not analyzed here.
\vspace{-3mm}

\subsection{Data Analysis}

\subsubsection{How we applied network analysis metrics to scanpaths.}

\begin{figure}[!ht]\centering\includegraphics[width=\columnwidth]{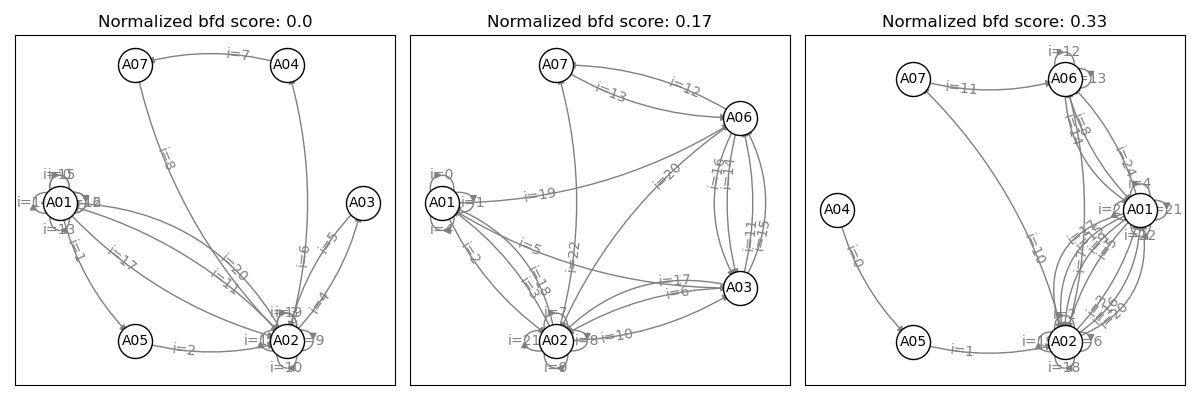}
\vspace{-7mm}

 \caption{Network representations of one participant's eye-gaze movements.}
 \label{fig:network_representation}
  \vspace{-6mm}
\end{figure}

Network Analysis is used to represent relational data in a network format (e.g., a graph) and apply network metrics and algorithms to the data. A first step in a network analysis project is to construct a network based on the analytical goals. To this end, we constructed ego networks (one per person per trial) by converting participants' eye-gaze movement data into network representations. In our data, nodes represent participants' focus on individual AOIs (Areas of Interest) as predefined by~\cite{borchers2023time} and edges represent movements or transitions (saccades) between AOIs. Given that prior research has often observed back-and-forth movements between nodes, we used multi-edge directed graphs weighted by cumulative frequency of eye-gaze movements to model these networks. The constructed networks represent the movement of eye-gazing across objects and may provide insights into cognitive processes during tasks. Figure~\ref{fig:network_representation} illustrates three network representations for one participant on subsequent study trials. Network representations capture transitions and the temporal order of movements between AOIs. 




\vspace{-7mm}

Using this definition and representation of eye-gaze networks, we applied mixed linear models to test for correlations between network metrics and measured variables. Figure~\ref{fig:network_representation} shows how network representations may correlate with students’ performance. The anomaly detection score (normalized BFD score, where BFD refers to "Befund" in German) refers to an adjusted percentage score, scaled from 0 to 1 (lowest to highest performance), which reflects students’ performance on the same OPT task that they repeated three times within the same semester. This allows us to form hypotheses, such as whether the number of edges in a network correlates with normalized BFD scores or not.

Based on a literature review~\cite{shiferaw2019review} of network analysis metrics applied to eye-gaze data such as shifts among AOIs in eye tracking data, shifts between geospatial location by pedestrians, or switches among different tools in virtual environments, we selected the following four categories of network analysis metrics: basic (node and edge count) measures, centrality measures, network-level structural measures, and entropy measures.

Basic measures refer to the number of nodes and edges per network. The number of nodes represents the AOIs a student has explored; the number of edges represents each student's total movements across AOIs. Existing research has shown that a decrease in the number of nodes may suggest higher learning efficiency~\cite{starke2018effect} and a positive correlation between the number of edges with the performance of 8th-grade students during problem solving~\cite{zhu2015exploratory}. 

Node centralities refer to a collection of metrics that capture different dimensions of the structural importance of individual nodes. Following existing research, we selected degree centrality~\cite{davalos2023identifying}, betweenness centrality~\cite{jayasinghe2017application}, closeness centrality~\cite{jayasinghe2017application}, eigenvector centrality~\cite{ryabinin2023eye}, and pagerank centrality~\cite{kakatkar2019marketing}. 

Structural measures tap characteristics of a network overall, such as the overall interconnectedness of nodes. For example, density, a measure of actual edges in a graph as a proportion of all possible edges, was reported to be positively related to learning outcomes~\cite{starke2017workflows}. Reciprocity, a measure of the ratio of back-and-forth movement, can be correlated with better outcomes when it represents integrating information between different sources or locations~\cite{li2022patterns}. Node connectivity is equal to the minimum number of nodes that must be removed to disconnect the parts of a network or render it a disconnected graph~\cite{cela2015social,esfahanian2013connectivity}. 

Entropy measures~\cite{krejtz2014entropy} are less frequently used than the other mentioned metrics. Prior work has shown that entropy measures are predictive of pilot training performance~\cite{diaz2019effects} and correlated with learning under time pressure in an ESL learning context ~\cite{jonsdottir2023effects}. The present research expands the use of entropy metrics to a longitudinal context. Stationary entropy quantifies aspects of the typical heat map representations of eye tracking by capturing the distribution of eye gazing across different AOIs. Transition entropy measures unpredictability or randomness in the sequence of transitions. Eye tracking research in medical education with intermediate learners and experts has shown that transition entropy decreases with expertise, as does stationary entropy~\cite{mozaffari2018evaluating}. 

\vspace{-5mm}

\subsubsection{Analysis of time series data and data visualization.}

When analyzing longitudinal data, methodologists emphasize the importance of inspecting patterns of change visually~\cite{singer2003applied} or quantitatively. In our data, visual inspection of time series plots of all network analysis metrics suggested 1) a general decreasing trend for each metric within each session (examining ten X-rays also known as OPTs), suggesting that eye-gaze patterns were becoming simpler, and 2) increasing trendings of metrics across sessions, indicating that eye-gaze patterns were reverting to more complex patterns, with a great deal of variability in the data. Thus, we decided to use time series analysis to deal with large within-person variability in network metrics at closely spaced measurement times~\cite{huang2017centrality}. To implement that, we used a k-means-based time series clustering method to analyze longitudinal eye-gaze data. The k-means-based approach encompasses a family of machine learning algorithms designed to group data points into distinct clusters based on similarity. Each cluster is represented by a “centroid,” which is the average position of all points in the group. The k-means-based approach has been widely used for both classification and clustering tasks, demonstrating reliable performance in analyzing behavioral data such as physical exercise and study time~\cite{chang2020analysis}. Since our objective is to test for latent classes within time series data, we used unsupervised k-means clustering algorithms to identify groups within the dataset. Specifically, we leverage the tslearn library in Python for time series clustering. Given that our dataset includes time series of varying lengths, we adopt the dynamic time warping (DTW) distance metric to compute averages across participants. DTW has been shown to provide more accurate similarity measurements for time series data than traditional Euclidean distance metrics.

To examine the relationships between network analysis metrics and participant performance, we use a generalized additive model (GAM) to fit a smooth curve to these non-linear data. This approach helps uncover temporal patterns in how network metrics relate to participant performance (i.e., detecting/annotating dental anomalies from the X-rays). GAM extends linear models to accommodate different data types, and facilitates the differentiation of fixed from random effects within a single model, providing flexibility and precision in analyzing complex relationships~\cite{hastie2017generalized}.

\vspace{-4mm}

\section{Results}

We first report the time series clustering of students' network representations and the performance on identifying anomalies in the X-rays for students in each cluster. We then present our findings from statistical analyses of the relationship between network metrics and students' performance and explore indicators for varying students' performance. 

\vspace{-3mm}

\subsection{Time Series Clustering of Students' Eye-gaze Network Metrics over Time}

\begin{figure}[!ht]
 \centering     
 
 \includegraphics[width=1\columnwidth]{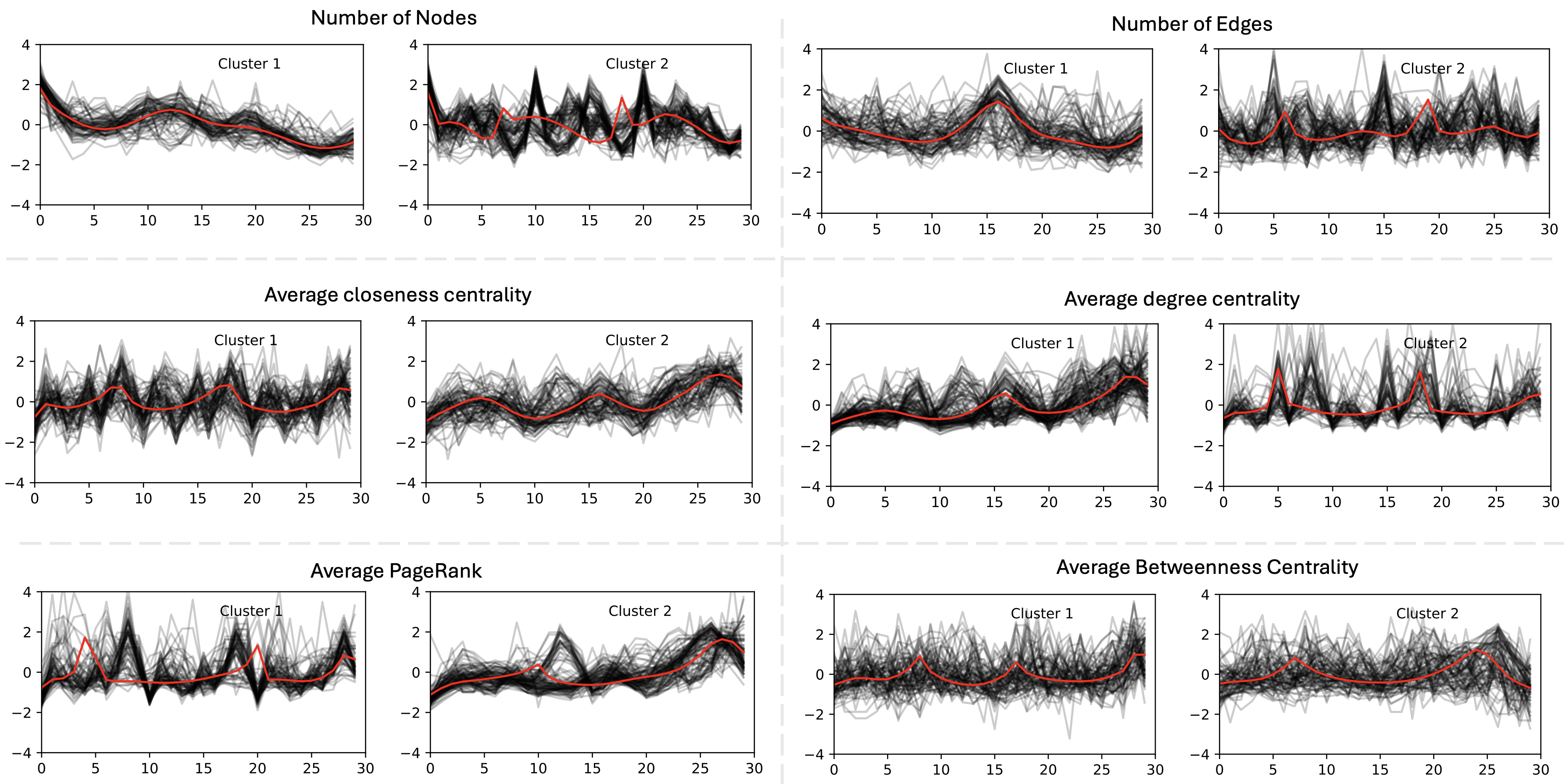}
 \includegraphics[width=1\columnwidth]{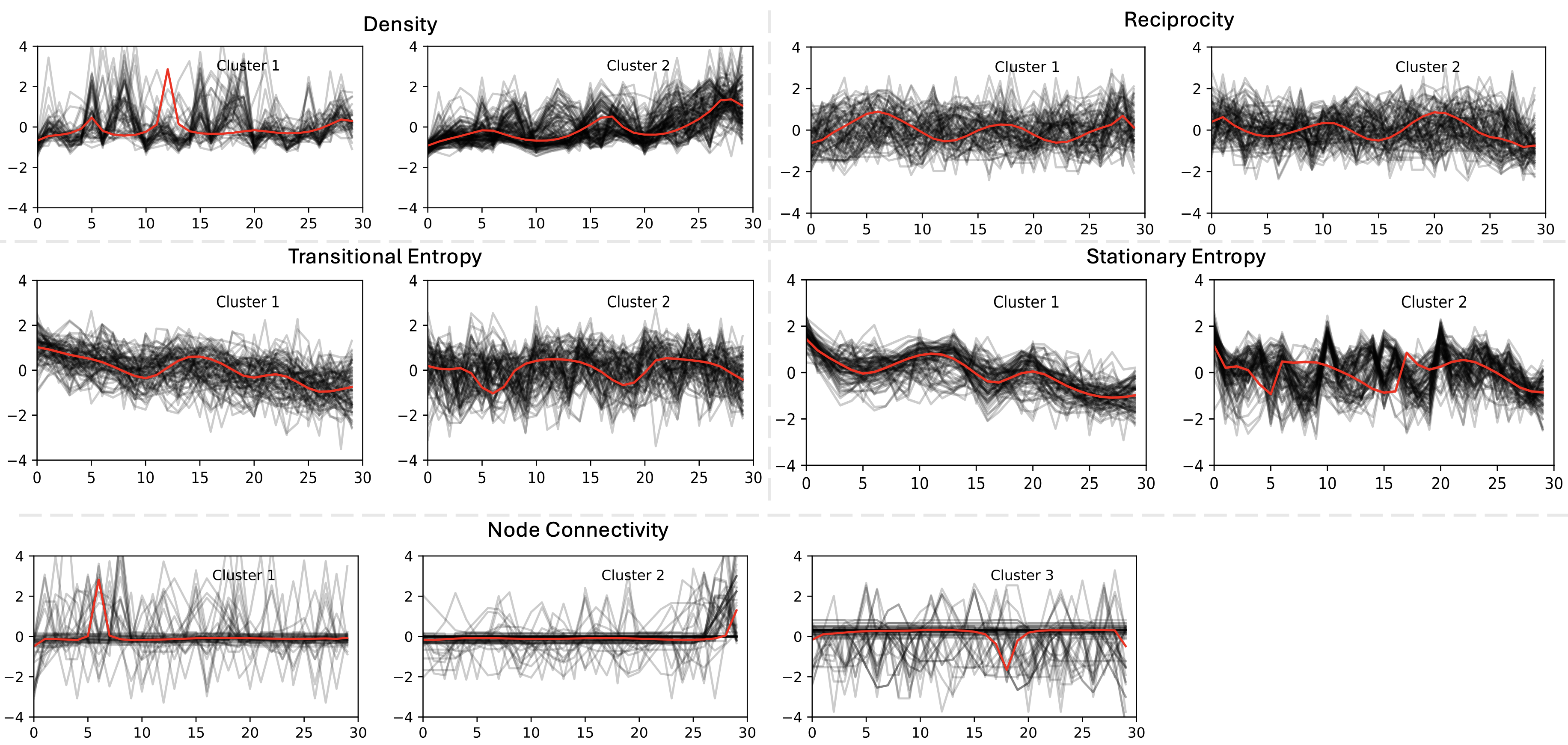}
\vspace{-7mm}

 \caption{K-means-based clustering of participants based on eleven network metrics.}
 \label{fig:clusterings}
  \vspace{-5mm}
\end{figure}

Figure~\ref{fig:clusterings} presents the results of k-means-based clustering of participants' network metric scores for eleven network metrics. Each subplot illustrates the temporal evolution of a specific network metrics for participants within each identified cluster. The x-axis represents observation moments over time, and the y-axis denotes the normalized values of the respective metric.

Thin black lines in each subplot depict the individual trajectories of participants within the cluster, showing variability in their behaviors over time. The red line is the cluster-level mean trend for the metric, providing a visualization of the general pattern for participants in each cluster on each metric.

The results show distinct patterns in network metrics across clusters. For example, metrics such as Node Connectivity display trends across the three clusters: Cluster 1 shows a pattern of decrease over time, Cluster 2 shows an increasing trend, and Cluster 3 shows a relative stable trend in Figure~\ref{fig:clusterings}. Cluster numbers are arbitrary and do not imply any ordinal relationship. Similarly, variations in the growth of metrics, e.g., for Reciprocity and Density, suggest potential differences in participants' diagnostic strategies or task engagement, which we will show to be related to performance in Section 5.3. Metrics such as the number of nodes and the number of edges demonstrate more similar trends between clusters, but still exhibit observable differences at certain time intervals, suggesting variations in learning processes.

\vspace{-3mm}

\subsection{BFD Performance Comparison across Time (ANOVA).}

Following the clustering of students based on network metrics calculated over networks representing the students' scan paths when visually inspecting X-rays, we compared their post-experiment performance across eleven metrics using an ANOVA (see  Table~\ref{tab:group_stat} for results). Significant differences were observed for node connectivity (\textit{F} = 4.205, \textit{p} = 0.015) and reciprocity (\textit{F} = 4.462, \textit{p} = 0.035), indicating that students with better performance may exhibit more sophisticated visual exploration strategies and deeper engagement with the task. Specifically, Cluster 1 for node connectivity showed a higher BFD score, and Cluster 2 showed a higher BFD for reciprocity, suggesting that these metrics may play a role in differentiating student performance across groups. However, there were no statistically significant differences between BFD scores among the clusters for the majority of the metrics we considered (see Table~\ref{tab:group_stat}).


Results suggest that while most network metrics do not vary significantly between clusters, i.e., both patterns of change are equally adaptive in terms of leading to higher BFD scores, structural features such as node connectivity and reciprocity may be important indicators of group differences in post-experiment performance. Further exploration of the correlations of these metrics with learning outcomes could provide deeper insights into their relevance in student clustering and performance outcomes.
\vspace{-5mm}
\begin{table}[htbp]
\caption{Comparison of mean BFD scores across groups. N-Mean-1 refers to the mean of the normalized value for the corresponding metric in Cluster 1, and BFD-1 refers to the mean BFD score for Cluster 1. The other columns follow the same rule.}
\label{tab:group_stat}
\begin{tabular}{|p{3cm}|l|l|l|l|l|l|l|l|}
\hline
\textbf{Metric}              & \textbf{N-Mean-1} & \textbf{N-Mean-2} & \textbf{N-Mean-3} & \textbf{BFD-1} & \textbf{BFD-2} & \textbf{BFD-3} & \textbf{f-stat} & \textbf{p-stat} \\ \hline
number of nodes            & -0.0006           & -0.0302      & na                & 0.397          & 0.408          & na             & 0.582           & 0.446           \\ \hline
number of edges            & -0.0097           & 0.0069            & na                & 0.397          & 0.409          & na             & 0.740           & 0.390           \\ \hline
avg degree centrality         & 0.0007            & 0.0040            & na                & 0.399          & 0.411          & na             & 0.614           & 0.433           \\ \hline
avg closeness centrality   & 0.0060            & 0.0014            & na                & 0.403          & 0.403          & na             & 0.002           & 0.960           \\ \hline
avg pagerank                & -0.0003           & 0.0125            & na                & 0.398          & 0.407          & na             & 0.385           & 0.535           \\ \hline
avg betweenness centrality & 0.0002            & 0.0170            & na                & 0.407          & 0.399          & na             & 0.287           & 0.592           \\ \hline
density                      & 0.0005            & 0.0040            & na                & 0.381          & 0.412          & na             & 3.723           & 0.054           \\ \hline
node connectivity           & -0.0040           & -0.0092      & 0.031          & 0.398          & 0.381          & 0.429          & 4.205           & 0.015*           \\ \hline
reciprocity                  & 0.0012            & 0.0078            & na                & 0.419          & 0.389          & na             & 4.462           & 0.035*           \\ \hline
stationary entropy                         & -0.0145           & -0.0022      & na                & 0.391          & 0.414          & na             & 2.421           & 0.120           \\ \hline
transition entropy                         & -0.0027           & -0.0003      & na                & 0.397          & 0.409          & na             & 0.674           & 0.412           \\ \hline
\end{tabular}
\end{table}


\subsection{Prediction Models of X-ray Reading Performance Based on Network Metrics}

We used regression (mixed linear model) to predict students' BFD scores from network metrics (see table~\ref{tab:regression} for results). The analysis included 3,425 observations nested within 165 participants from the original dataset, with a mean group size of 20.8 (range: 10–50). The model successfully converged using the Restricted Maximum Likelihood (REML) estimation method (Log-Likelihood = -416.1442, Scale = 0.0617) using the \textit{statsmodels} package in Python.

From the fixed effects portion of the model, we can see that the OPT Task Ordered Index ($\beta$ = -0.030, \textit{SE} = 0.014, \textit{z} = -2.209, \textit{p} = 0.027) and transition entropy ($\beta$ = -0.065, \textit{SE} = 0.028, \textit{z} = -2.291, \textit{p} = 0.022) are significant negative predictors of the normalized BFD score, suggesting that students with better performance tend to exhibit less random scanning paths when completing tasks. Conversely, the number of nodes ($\beta$ = 0.015, \textit{SE} = 0.006, \textit{z} = 2.761, \textit{p} = 0.006),\textit{} number of edges ($\beta$ = 0.002, \textit{SE} = 0.001, \textit{z} = 3.398, \textit{p} = 0.001), and average PageRank ($\beta$ = 0.701, \textit{SE} = 0.192, \textit{z} = 3.651, \textit{p} < 0.001) were significant positive predictors, suggesting that students with better performance tend to explore more nodes and connections, and place greater visual emphasis on conceptually important AOIs. Other potential predictors, including average degree centrality, density, and reciprocity, were not significant (all \textit{p} > 0.05).



\begin{table}[htbp]
\caption{Predicting students' BFD score using network metrics}
\label{tab:regression}
\begin{tabular}{l|l|l|llll}
\cline{2-7}
                                                              & \textbf{Coef.} & \textbf{Std.Err.} & \multicolumn{1}{l|}{\textbf{z}} & \multicolumn{1}{l|}{\textbf{P\textgreater{}|z|}} & \multicolumn{1}{l|}{\textbf{{[}0.025}} & \multicolumn{1}{l|}{\textbf{0.975{]}}} \\ \hline
\multicolumn{1}{|l|}{\textbf{Intercept}}                      & 0.203          & 0.129             & \multicolumn{1}{l|}{1.57}       & \multicolumn{1}{l|}{0.116}                       & \multicolumn{1}{l|}{-0.05}             & \multicolumn{1}{l|}{0.457}             \\ \hline
\multicolumn{1}{|l|}{\textbf{Time}}            & -0.03          & 0.014             & \multicolumn{1}{l|}{-2.209}     & \multicolumn{1}{l|}{0.027}                       & \multicolumn{1}{l|}{-0.057}            & \multicolumn{1}{l|}{-0.003}            \\ \hline
\multicolumn{1}{|l|}{\textbf{Stationary Entropy}}             & 0.042          & 0.022             & \multicolumn{1}{l|}{1.93}       & \multicolumn{1}{l|}{0.054}                       & \multicolumn{1}{l|}{-0.001}            & \multicolumn{1}{l|}{0.086}             \\ \hline
\multicolumn{1}{|l|}{\textbf{Transition Entropy}}             & -0.065         & 0.028             & \multicolumn{1}{l|}{-2.291}     & \multicolumn{1}{l|}{0.022*}                       & \multicolumn{1}{l|}{-0.12}             & \multicolumn{1}{l|}{-0.009}            \\ \hline
\multicolumn{1}{|l|}{\textbf{Number of Nodes}}                & 0.015          & 0.006             & \multicolumn{1}{l|}{2.761}      & \multicolumn{1}{l|}{0.006*}                       & \multicolumn{1}{l|}{0.004}             & \multicolumn{1}{l|}{0.026}             \\ \hline
\multicolumn{1}{|l|}{\textbf{Number of Edges}}                & 0.002          & 0.001             & \multicolumn{1}{l|}{3.398}      & \multicolumn{1}{l|}{0.001*}                       & \multicolumn{1}{l|}{0.001}             & \multicolumn{1}{l|}{0.003}             \\ \hline
\multicolumn{1}{|l|}{\textbf{Average Degree Centrality}}      & -0.256         & 0.164             & \multicolumn{1}{l|}{-1.563}     & \multicolumn{1}{l|}{0.118}                       & \multicolumn{1}{l|}{-0.577}            & \multicolumn{1}{l|}{0.065}             \\ \hline
\multicolumn{1}{|l|}{\textbf{Average Betweenness Centrality}} & 0.115          & 0.069             & \multicolumn{1}{l|}{1.669}      & \multicolumn{1}{l|}{0.095}                       & \multicolumn{1}{l|}{-0.02}             & \multicolumn{1}{l|}{0.251}             \\ \hline
\multicolumn{1}{|l|}{\textbf{Average Closeness Centrality}}   & 0.082          & 0.074             & \multicolumn{1}{l|}{1.109}      & \multicolumn{1}{l|}{0.267}                       & \multicolumn{1}{l|}{-0.063}            & \multicolumn{1}{l|}{0.226}             \\ \hline
\multicolumn{1}{|l|}{\textbf{Average PageRank}}               & 0.701          & 0.192             & \multicolumn{1}{l|}{3.651}      & \multicolumn{1}{l|}{0.001*}                           & \multicolumn{1}{l|}{0.324}             & \multicolumn{1}{l|}{1.077}             \\ \hline
\multicolumn{1}{|l|}{\textbf{Density}}                        & 0.509          & 0.324             & \multicolumn{1}{l|}{1.57}       & \multicolumn{1}{l|}{0.117}                       & \multicolumn{1}{l|}{-0.127}            & \multicolumn{1}{l|}{1.145}             \\ \hline
\multicolumn{1}{|l|}{\textbf{Reciprocity}}                    & 0.069          & 0.045             & \multicolumn{1}{l|}{1.556}      & \multicolumn{1}{l|}{0.12}                        & \multicolumn{1}{l|}{-0.018}            & \multicolumn{1}{l|}{0.157}             \\ \hline
\multicolumn{1}{|l|}{\textbf{Node Connectivity}}              & -0.004         & 0.003             & \multicolumn{1}{l|}{-1.428}     & \multicolumn{1}{l|}{0.153}                       & \multicolumn{1}{l|}{-0.009}            & \multicolumn{1}{l|}{0.001}             \\ \hline
\multicolumn{1}{|l|}{\textbf{Participant Var}}                & 1.293          & 0.94              &                                 &                                                  &                                        &                                        \\ \cline{1-3}
\multicolumn{1}{|l|}{\textbf{Participant x Semester Cov}}     & -0.181         & 0.131             &                                 &                                                  &                                        &                                        \\ \cline{1-3}
\multicolumn{1}{|l|}{\textbf{Semester Var}}                   & 0.026          & 0.018             &                                 &                                                  &                                        &                                        \\ \cline{1-3}
\end{tabular}
\vspace{-5mm}

\end{table}
\vspace{-4mm}

\section{Discussion}

\subsection{Is there a Relationship between network metrics and earning Performance?}

We observe mostly linear changes in students’ BFD scores and eye-gaze patterns over time. Among the eleven metrics we examined, transition entropy, number of nodes and edges, and average PageRank were significantly correlated with students’ BFD scores. Changes in network metrics, including the number of nodes and edges, and average PageRank, suggest that participants are transitioning from an intermediate to an expert skill level in reading X-rays. The values of network metrics did not suggest that participants were following the recommended top-left-to-top-right, bottom-right-to-bottom-left inspection patterns~\cite{cosson2020interpreting}.

Most of the participants in the dataset we reused could be clustered into one of two or three categories for each of the eleven network metrics, indicating different approaches to the task. However, performance mostly did not differ between clusters. Only node connectivity and reciprocity show significant differences in BFD scores across clusters. As reciprocity measures how likely nodes are interconnected, high reciprocity may be a signal of confusion in individual eye-gaze networks, which aligns with our observation that high reciprocity is associated with low student BFD scores. Higher node connectivity may indicate a higher modularity in the network, suggesting learners tend to organize gaze patterns into well-defined sub-areas of the display. 

Transition entropy is low for the very efficient eye-gaze patterns of experts, and here we saw how entropy decreases with instruction. That is, the participants showed more random, exploratory scanpaths in earlier sessions and more directed scanpaths in later sessions, indicative of a shift toward having acquired visual expertise.

\vspace{-3mm}

\subsection{Enabling Personalized Feedback and Learning}

As our study finds that certain gaze-based, network analysis metrics are correlated with X-ray reading performance, these metrics could also be adapted to actively intervene in students’ learning processes. Without explicit external assessments, these metrics could provide self-assessment feedback to students during the learning process. Additionally, at different stages of learning, these network-based metrics could be used to actively track students’ learning progress and enable personalized interventions when necessary. Although some studies have shown that gaze data can be used for AI-generated adaptive interventions to improve student learning outcomes~\cite{santhosh2024gaze}, the integration of such interventions into real-time systems remains in its early stages, primarily due to the lack of high-quality data collected through non-intrusive and distraction-free methods. Our clustering and tracking of metrics over time also indicate that network metrics, such as transition entropy, number of nodes, number of edges, and average PageRank, significantly correlate with students’ BFD scores. For domain-specific visual tasks, similar approaches could be applied to behavioral data to develop optimized or customized models.

From an instructor's perspective, such network metrics can also provide complementary evidence that may be invisible through traditional assessments, such as tests, to track students’ learning progress in visually oriented tasks. Even outside of the context of a specific assignment, changes in students’ visual behaviors can serve as a valuable source for evaluating learning progress.

Another consideration for personalized feedback is model transparency. Compared to deep learning-based approaches, our method has lower computational complexity and greater transparency. Network metrics can be traced back to visual network representations, providing an explainable link between NA metrics, BFD scores, and students’ behaviors.

\vspace{-4mm}

\section{Limitations and Future Work}

As far as we know, this study is the first work to apply network analytic metrics to relationally represented eye-gaze data in a longitudinal setting. There are some improvements to be made in future work. First, we were not able to use the whole dataset due to model restrictions on analyzing varying-length time-series data. Future work should explore a neural temporal encoder approach to analyze the whole dataset. Second, lacking information about the difficulty level of the X-ray tasks may introduce bias into the analyses. Subsequent work should apply IRT scaling to create weights for different OPT tasks. Finally, the expertise level of participants is unknown to us, which makes it difficult to make strong claims in our interpretation of what the changes in network metrics mean. The data we used were anonymized in the original dataset and hence cannot be re-identified. More developmental eye-tracking studies are needed to understand the progression from novice to intermediate to expert. 
\vspace{-4mm}

\section{Conclusions}

We demonstrated that network metrics calculated on eye-gazing data, such as transition entropy, number of nodes or edges, and average PageRank, are correlated with X-ray reading performance and track the development of visual expertise in dentistry students. Observed changes in these metrics suggest a shift from intermediate to expert-like search patterns, while clustering reveals distinct developmental trajectories. Our findings highlight the potential of using gaze-based metrics for real-time, self-assessment feedback and personalized learning interventions in visual learning tasks.

\section*{Acknowledgment}

Pingjing Yang is working in the Social Computing Lab, led by Professor Jana Diesner, at the University of Illinois at Urbana-Champaign. This work is conducted under the mentorship of Professor Jennifer Cromley in the College of Education at the University of Illinois at Urbana-Champaign. The research was supported with funds from the US National Science Foundation under Award No. 2225298 to the University of Illinois at Urbana-Champaign. Any opinions, findings, and conclusions or recommendations expressed in this material are those of the authors and do not necessarily reflect the views of the National Science Foundation.

\bibliographystyle{splncs04}
\bibliography{references}

\begin{thebibliography}{10}
\providecommand{\url}[1]{\texttt{#1}}
\providecommand{\urlprefix}{URL }
\providecommand{\doi}[1]{https://doi.org/#1}

\bibitem{ashraf2018eye}
Ashraf, H., Sodergren, M.H., Merali, N., Mylonas, G., Singh, H., Darzi, A.:
  Eye-tracking technology in medical education: A systematic review. Medical
  teacher  \textbf{40}(1),  62--69 (2018)

\bibitem{bernard2016analyzing}
Bernard, H.R., Wutich, A., Ryan, G.W.: Analyzing qualitative data: Systematic
  approaches. SAGE publications (2016)

\bibitem{black1999predicting}
Black, M.M., Krishnakumar, A.: Predicting longitudinal growth curves of height
  and weight using ecological factors for children with and without early
  growth deficiency. The Journal of nutrition  \textbf{129}(2),  539S--543S
  (1999)

\bibitem{borchers2023time}
Borchers, C., Eder, T.F., Richter, J., Keutel, C., Huettig, F., Scheiter, K.: A
  time slice analysis of dentistry students’ visual search strategies and
  pupil dilation during diagnosing radiographs. PLoS One  \textbf{18}(6),
  e0283376 (2023)

\bibitem{castner2022gaze}
Castner, N., Umlauf, B., Kastrati, A., P{\l}omecka, M.B., Schaefer, W.,
  Kasneci, E., Bylinskii, Z.: A gaze-based study design to explore how
  competency evolves during a photo manipulation task. In: 2022 Symposium on
  Eye Tracking Research and Applications. pp.~1--3 (2022)

\bibitem{cela2015social}
Cela, K.L., Sicilia, M.{\'A}., S{\'a}nchez, S.: Social network analysis in
  e-learning environments: A preliminary systematic review. Educational
  psychology review  \textbf{27},  219--246 (2015)

\bibitem{chang2020analysis}
Chang, W., Ji, X., Liu, Y., Xiao, Y., Chen, B., Liu, H., Zhou, S.: Analysis of
  university students’ behavior based on a fusion k-means clustering
  algorithm. Applied Sciences  \textbf{10}(18), ~6566 (2020)

\bibitem{chanijani2016entropy}
Chanijani, S.S.M., Klein, P., Bukhari, S.S., Kuhn, J., Dengel, A.: Entropy
  based transition analysis of eye movement on physics representational
  competence. In: Proceedings of the 2016 ACM international joint conference on
  pervasive and ubiquitous computing: adjunct. pp. 1027--1034 (2016)

\bibitem{cosson2020interpreting}
Cosson, J.: Interpreting an orthopantomogram. Australian Journal of General
  Practice  \textbf{49}(9),  550--555 (2020)

\bibitem{davalos2023identifying}
Davalos, E., Vatral, C., Cohn, C., Horn~Fonteles, J., Biswas, G., Mohammed, N.,
  Lee, M., Levin, D.: Identifying gaze behavior evolution via temporal
  fully-weighted scanpath graphs. In: LAK23: 13th International Learning
  Analytics and Knowledge Conference. pp. 476--487 (2023)

\bibitem{dhengre2024investigating}
Dhengre, S., Nam, H., Helm, M., Rothrock, L.: Investigating effect of
  standardized total body skin examination using sequence-networks. Applied
  Ergonomics  \textbf{116},  104219 (2024)

\bibitem{diaz2019effects}
Diaz-Piedra, C., Rieiro, H., Cherino, A., Fuentes, L.J., Catena, A., Di~Stasi,
  L.L.: The effects of flight complexity on gaze entropy: An experimental study
  with fighter pilots. Applied ergonomics  \textbf{77},  92--99 (2019)

\bibitem{eder2021support}
Eder, T.F., Richter, J., Scheiter, K., Keutel, C., Castner, N., Kasneci, E.,
  Huettig, F.: How to support dental students in reading radiographs: effects
  of a gaze-based compare-and-contrast intervention. Advances in Health
  Sciences Education  \textbf{26},  159--181 (2021)

\bibitem{eder2021comparing}
Eder, T.F., Richter, J., Scheiter, K., Huettig, F., Keutel, C.: Comparing
  radiographs with signaling improves anomaly detection of dental students: An
  eye-tracking study. Applied Cognitive Psychology  \textbf{35}(4),  909--923
  (2021)

\bibitem{eder2022see}
Eder, T.F., Scheiter, K., Richter, J., Keutel, C., H{\"u}ttig, F.: I see
  something you do not: Eye movement modelling examples do not improve anomaly
  detection in interpreting medical images. Journal of Computer Assisted
  Learning  \textbf{38}(2),  379--391 (2022)

\bibitem{esfahanian2013connectivity}
Esfahanian, A.H.: Connectivity algorithms. Topics in structural graph theory
  pp. 268--281 (2013)

\bibitem{grub2022professional}
Grub, A.S., Biermann, A., Lewalter, D., Br{\"u}nken, R.: Professional vision
  and the compensatory effect of a minimal instructional intervention: a
  quasi-experimental eye-tracking study with novice and expert teachers. In:
  Frontiers in Education. vol.~7, p. 890690. Frontiers Media SA (2022)

\bibitem{hastie2017generalized}
Hastie, T.J.: Generalized additive models. Statistical models in S pp. 249--307
  (2017)

\bibitem{hostinar2014cortisol}
Hostinar, C.E., McQuillan, M.T., Mirous, H.J., Grant, K.E., Adam, E.K.:
  Cortisol responses to a group public speaking task for adolescents:
  Variations by age, gender, and race. Psychoneuroendocrinology  \textbf{50},
  155--166 (2014)

\bibitem{hua2022effect}
Hua, Y., Fu, S., Lu, Y.: The effect of pilots’ expertise on eye movement and
  scan patterns during simulated flight tasks. In: International Conference on
  Human-Computer Interaction. pp. 290--299. Springer (2022)

\bibitem{huang2017centrality}
Huang, Q., Zhao, C., Zhang, X., Wang, X., Yi, D.: Centrality measures in
  temporal networks with time series analysis. Europhysics Letters
  \textbf{118}(3),  36001 (2017)

\bibitem{ifenthaler2011mystery}
Ifenthaler, D., Masduki, I., Seel, N.M.: The mystery of cognitive structure and
  how we can detect it: tracking the development of cognitive structures over
  time. Instructional Science  \textbf{39},  41--61 (2011)

\bibitem{jayasinghe2017application}
Jayasinghe, A., Sano, K., Rattanaporn, K.: Application for developing
  countries: Estimating trip attraction in urban zones based on centrality.
  Journal of Traffic and Transportation Engineering (English Edition)
  \textbf{4}(5),  464--476 (2017)

\bibitem{jonsdottir2023effects}
J{\'o}nsd{\'o}ttir, A.A., Kang, Z., Sun, T., Mandal, S., Kim, J.E.: The effects
  of language barriers and time constraints on online learning performance: An
  eye-tracking study. Human Factors  \textbf{65}(5),  779--791 (2023)

\bibitem{kakatkar2019marketing}
Kakatkar, C., Spann, M.: Marketing analytics using anonymized and fragmented
  tracking data. International Journal of Research in Marketing
  \textbf{36}(1),  117--136 (2019)

\bibitem{krejtz2014entropy}
Krejtz, K., Szmidt, T., Duchowski, A.T., Krejtz, I.: Entropy-based statistical
  analysis of eye movement transitions. In: Proceedings of the Symposium on Eye
  Tracking Research and Applications. pp. 159--166 (2014)

\bibitem{leng2024train}
Leng, X., Wang, F., Mayer, R.E., Zhao, T.: How to train students to engage in
  text-picture integration for multimedia lessons. British Journal of
  Educational Technology  \textbf{55}(3),  1167--1188 (2024)

\bibitem{li2022patterns}
Li, S., P{\"o}ys{\"a}-Tarhonen, J., H{\"a}kkinen, P.: Patterns of action
  transitions in online collaborative problem solving: A network analysis
  approach. International Journal of Computer-Supported Collaborative Learning
  \textbf{17}(2),  191--223 (2022)

\bibitem{loyola2014characterizing}
Loyola, P., Vel{\'a}squez, J.D.: Characterizing web user visual gaze patterns:
  A graph theory inspired approach. In: Brain Informatics and Health:
  International Conference, BIH 2014, Warsaw, Poland, August 11-14, 2014,
  Proceedings. pp. 586--594. Springer (2014)

\bibitem{ma2023eye}
Ma, X., Liu, Y., Clariana, R., Gu, C., Li, P.: From eye movements to scanpath
  networks: A method for studying individual differences in expository text
  reading. Behavior research methods  \textbf{55}(2),  730--750 (2023)

\bibitem{mason2013fourth}
Mason, L., Tornatora, M.C., Pluchino, P.: Do fourth graders integrate text and
  picture in processing and learning from an illustrated science text? evidence
  from eye-movement patterns. Computers \& Education  \textbf{60}(1),  95--109
  (2013)

\bibitem{mozaffari2018evaluating}
Mozaffari, S., Klein, P., Viiri, J., Ahmed, S., Kuhn, J., Dengel, A.:
  Evaluating similarity measures for gaze patterns in the context of
  representational competence in physics education. In: Proceedings of the 2018
  ACM Symposium on Eye Tracking Research \& Applications. pp.~1--5 (2018)

\bibitem{pianta2008classroom}
Pianta, R.C., Belsky, J., Vandergrift, N., Houts, R., Morrison, F.J.: Classroom
  effects on children’s achievement trajectories in elementary school.
  American educational research journal  \textbf{45}(2),  365--397 (2008)

\bibitem{richter2020massed}
Richter, J., Scheiter, K., Eder, T.F., Huettig, F., Keutel, C.: How massed
  practice improves visual expertise in reading panoramic radiographs in dental
  students: An eye tracking study. PloS one  \textbf{15}(12),  e0243060 (2020)

\bibitem{rotgans2017interest}
Rotgans, J.I., Schmidt, H.G.: Interest development: Arousing situational
  interest affects the growth trajectory of individual interest. Contemporary
  Educational Psychology  \textbf{49},  175--184 (2017)

\bibitem{ryabinin2023eye}
Ryabinin, K., Erofeeva, E., Guseva, K.: Eye tracking data mining based on fuzzy
  sets of fixations. In: Fuzzy Systems and Data Mining IX, pp. 11--19. IOS
  Press (2023)

\bibitem{santhosh2024gaze}
Santhosh, J., Dengel, A., Ishimaru, S.: Gaze-driven adaptive learning system
  with chatgpt-generated summaries. IEEE Access  \textbf{12},  173714--173733
  (2024)

\bibitem{schieber2008visual}
Schieber, F., Gilland, J.: Visual entropy metric reveals differences in
  drivers' eye gaze complexity across variations in age and subsidiary task
  load. In: Proceedings of the Human Factors and Ergonomics Society Annual
  Meeting. vol.~52, pp. 1883--1887. SAGE Publications Sage CA: Los Angeles, CA
  (2008)

\bibitem{shiferaw2019review}
Shiferaw, B., Downey, L., Crewther, D.: A review of gaze entropy as a measure
  of visual scanning efficiency. Neuroscience \& Biobehavioral Reviews
  \textbf{96},  353--366 (2019)

\bibitem{silva2022differences}
Silva, A.F., Afonso, J., Sampaio, A., Pimenta, N., Lima, R.F., Castro, H.d.O.,
  Ramirez-Campillo, R., Teoldo, I., Sarmento, H., Gonzalez~Fernandez, F.,
  et~al.: Differences in visual search behavior between expert and novice team
  sports athletes: A systematic review with meta-analysis. Frontiers in
  psychology  \textbf{13},  1001066 (2022)

\bibitem{singer2003applied}
Singer, J.D., Willett, J.B.: Applied Longitudinal Data Analysis: Modeling
  Change and Event Occurrence. Oxford University Press (05 2003).
  \doi{10.1093/acprof:oso/9780195152968.001.0001}

\bibitem{starke2018effect}
Starke, S.D., Baber, C.: The effect of four user interface concepts on visual
  scan pattern similarity and information foraging in a complex decision making
  task. Applied Ergonomics  \textbf{70},  6--17 (2018)

\bibitem{starke2017workflows}
Starke, S.D., Baber, C., Cooke, N.J., Howes, A.: Workflows and individual
  differences during visually guided routine tasks in a road traffic management
  control room. Applied ergonomics  \textbf{61},  79--89 (2017)

\bibitem{wickrama2021higher}
Wickrama, K., Lee, T.K., O’Neal, C.W., Lorenz, F.: Higher-order growth curves
  and mixture modeling with Mplus: A practical guide. Routledge (2021)

\bibitem{zhu2015exploratory}
Zhu, M., Feng, G.: An exploratory study using social network analysis to model
  eye movements in mathematics problem solving. In: Proceedings of the Fifth
  International Conference on Learning Analytics and Knowledge. pp. 383--387
  (2015)

\bibitem{ziv2016gaze}
Ziv, G.: Gaze behavior and visual attention: A review of eye tracking studies
  in aviation. The International Journal of Aviation Psychology
  \textbf{26}(3-4),  75--104 (2016)

\end{thebibliography}

\end{document}